\documentclass[prc,twocolumn,amsmath,letterpaper]{revtex4}
\usepackage{graphicx}
\usepackage{verbatim}

\begin{document}
\title{Universality of Decay out of Superdeformed Bands in the 190 Mass
  Region} \author{D. M. Cardamone} \affiliation{Physics
  Department, Simon Fraser University, Burnaby, BC V5A 1S6, Canada} \author{B.
  R. Barrett} \author{C. A. Stafford} \affiliation{Physics Department,
  University of Arizona, Tucson, AZ 85721, USA}

\begin{abstract}
  Superdeformed nuclei in the 190 mass region exhibit a striking universality
  in their decay-out profiles. 
We show that this universality can be explained
  in the two-level model of superdeformed decay as 
related
to 
the
 strong separation of energy scales: a
  higher scale related to the nuclear interactions, and a lower scale caused by
  electromagnetic decay. 
Decay-out can only occur when separate conditions in both energy regimes are
satisfied, strongly limiting the collective degrees of freedom available to the decaying nucleus.
Furthermore, we present the results of the
  two-level model for all decays for which sufficient data are known,
  including statistical extraction of the matrix element for tunneling through
  the potential barrier.
\end{abstract}

\maketitle

\section{Introduction}
It is well known that, for a major--to--minor axis ratio of about 2, a new set
of shell closures and magic numbers occurs in many nuclei. Such superdeformed
(SD) states are one of the most striking predictions of the shell model \cite{aberg90}. High
electric quadrupole moments and small centrifugal stretching mark these states
as fundamentally different from their normally deformed (ND) isomers \cite{twin90}. This
contrast has stimulated an abundance of experimental and theoretical studies,
yet several pressing questions persist \cite{aberg90,twin90,nolan88}.
Of these, perhaps the most interesting
is the mechanism by which SD bands decay.

\begin{figure}
  \includegraphics[angle=-90,width=\columnwidth,keepaspectratio=true]{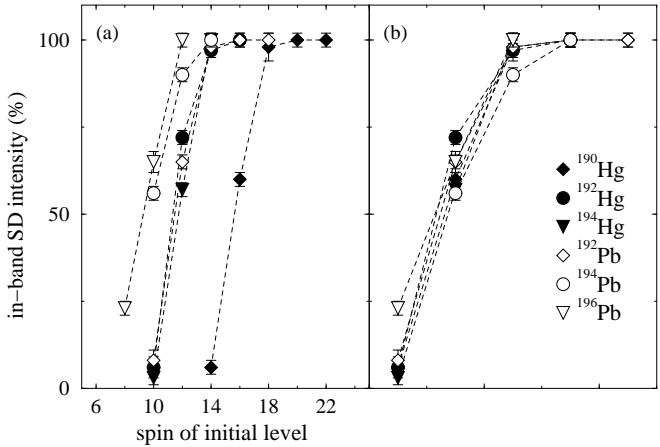}
  \caption{(a) Decay profiles of several SD bands near $A\approx 190$. Note
    how suddenly each decay-out occurs. (b) The profiles of (a), but shifted in
    angular momentum so that the leftmost points, the last point in which the
    SD band is experimentally observed to retain any strength, are aligned.
    In addition to the abruptness of their decay, the profiles are seen to exhibit a universal behavior. Both graphs are from Ref.\
    \cite{wilson05a}.}
  \label{profiles}
\end{figure}

After their formation at high angular momentum, typically via heavy ion
collisions, these nuclei decay to the yrast SD rotational band, and then
uniformly down that band by $E2$ transitions. The SD bands are observed
\begin{figure}[ht!]
  \includegraphics[keepaspectratio=false,width=\columnwidth,height=.5\columnwidth]{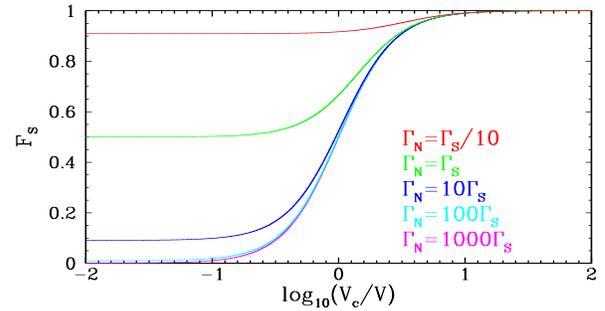}

  \caption{(Color online) SD branching ratio $F_S$, calculated in the two-level model and
    showing the onset of universality. 
    Universality of the decay
    profiles arises from the fact that interband decay is nearly forbidden
    until conditions are favorable in both energy regimes, at which point SD strength
    vanishes quickly. The sudden transition from intraband to
    interband decay occurs when both
    $\Gamma_N\gtrsim\Gamma_S$ and $V\gtrsim V_c$, where the critical tunneling
    matrix element $V_c$ is given by Eq.~(\ref{Vc}). When both conditions are
    satisfied, the curves are thus nearly identical, 
    giving rise to the observed universality of decay profiles.}
  \label{fig:results}
\end{figure}
to retain their strength through many states, even after they are no longer
yrast, with negligible losses. Then, quite suddenly, the SD band loses almost
all of its strength over just one or two states [see Fig.~\ref{profiles}(a)],
although the nucleus is still well above the SD bandhead. After a series of
statistical decays through unrelated states, the nuclei continue via
$E1$-dominated decays in an ND rotational band \cite{twin90,nolan88}.

By far the most SD decays have been observed in the ``classic'' 190 mass
region. Recently, Wilson and collaborators \cite{wilson05a} demonstrated a
striking 
feature of SD decay in this region: the decay profiles, when corrected for
differing angular momenta, are nearly identical [Fig.~\ref{profiles}(b)]. This
\emph{universality} of decays is to be contrasted with mere
\emph{abruptness}, a feature which has been acknowledged for some time.
Indeed, it
was noted as early as Ref.~\cite{shimizu93} that a purely statistical model
would be insufficient to explain true universality, i.e. strong consistency
between different decay profiles. Likewise, a chaos-assisted phenomenon 
cannot, of itself, generate universality.

The purpose of this article is to 
demonstrate that within the two-level model of SD decay-out, the decay
profile is universal.
In this model, the branching ratio is completely determined
by four parameters: the detuning $\Delta\equiv\varepsilon_N-\varepsilon_S$, a tunneling matrix element
$V$, and electromagnetically induced broadenings for the SD and ND wells, $\Gamma_S$ and
$\Gamma_N$, respectively. Because nuclear forces are very much stronger than
electromagnetic, there is a strong separation of scales
$V,\Delta\gg\Gamma_S,\Gamma_N$. We show that the decay-out only
occurs when conditions have become favorable in both energy regimes.
As a consequence, decay occurs not only very suddenly, but also in a limited region of the
  model's full parameter space. Within this subspace, branching ratios are highly
  insensitive to changes in the parameters, resulting in the universality observed in experiment.

Figure \ref{fig:results} illustrates these findings. It depicts the calculated value
of the in-band branching ratio $F_S$ for various ratios
$\Gamma_S/\Gamma_N$.
$V_c$, a simple function of $\Delta$ and $\Gamma_S/\Gamma_N$ in the two-level
model, 
sets the scale 
$V$ must achieve to allow decay. 
Universality is evident
from the figure: if either
$\Gamma_S/\Gamma_N$ or  $V_c/V$ is too large, no decay can occur.
On the other hand, as both cross critical values, 
$F_S$ suddenly vanishes. Furthermore, the curves converge in this, the
decay-allowed limit, resulting in a universal profile in good agreement with the experimental results of
Fig.~\ref{profiles}.
However, the connection between this theoretical universality and the 
experimentally observed universality still needs to be made, because the 
abscissas of the two plots (Figs. 1 and 2) are not the same.  This 
connection will require a theoretical understanding of the angular 
momentum dependence of $V$, which in turn requires a more detailed 
understanding of the barrier between the two wells.

The outline of the paper is as follows. We first briefly review the
two-level model in Section II. Section III presents the model's
results, on which Fig.~\ref{fig:results} is based. Section III further includes a
numerical analysis of all SD decays for which sufficient data are
known. Section IV gives our conclusions.

\section{Two-level model of SD decay}
\label{2-level}
Theoretical efforts to describe the SD decay process have centered on a potential
function of both nuclear quadrupole deformation and angular momentum. Vigezzi
and collaborators noted early on that a double well in the deformation more
accurately models the experimental data than any alternative \cite{vigezzi90a}.
In this picture, the shape of the tunnel barrier and the two
wells varies as the nucleus sheds angular momentum, and the states of the ND
and SD wells are broadened by their respective couplings to the
electromagnetic field. Shortly thereafter, Khoo \emph{et al.}\ \cite{khoo93a} conclusively
demonstrated that these electromagnetic widths must be much less than the
inter-level spacings in each well. The most appropriate
picture for SD decay is thus found to be two sets of discrete, slightly
broadened states, connected by matrix elements to tunnel through the barrier.

\begin{figure}
  \centerline{\includegraphics[keepaspectratio=true,width=.8\columnwidth]{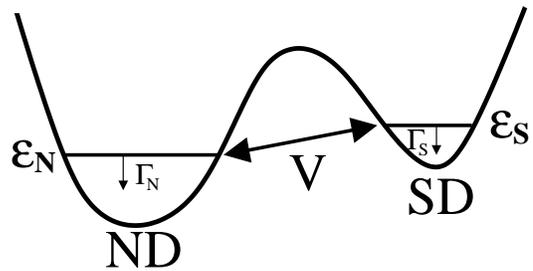}}
  \caption{Schematic diagram of the two-level model of SD decay. In each
    well of the double-well potential,
    only one level is kept. $\varepsilon_N$ and
    $\varepsilon_S$ are the unperturbed energies of the isolated ND and SD states, which are
    connected by a tunneling matrix element $V$. The two states have
    electromagnetic decay rates $\Gamma_N/\hbar$ and $\Gamma_S/\hbar$,
    respectively.}
  \label{2well}
\end{figure}

The two-level model for SD decay \cite{SB} is given by keeping only one level in each
well, the decaying SD level and the ND level with nearest energy (see
Fig.~\ref{2well}). Since the role of additional ND levels is
principally to steal decay strength from the first \cite{CSBPRL,dzyublik03a}, it
is now 
well established that going beyond this level of approximation is not useful
for most heavy-nuclei SD decays.

\subsection{Green's Function Description of SD Decay}
The Hamiltonian of the two-level model is a sum of three terms:
$H=H_W+H_T+H_D$. In the basis of the two isolated levels, the first term
is diagonal:
\begin{equation}
  H_W=\left(
\begin{array}{cc}
\varepsilon_S & 0\\
0 & \varepsilon_N
\end{array}
\right),
\end{equation}
where $\varepsilon_i$ is the energy of state $i$. $H_W$ generates time evolution within each well: its related
retarded Green's function is $G_W(E)=\left(E-H_W+i0^+\right)^{-1}$.

The second term,
\begin{equation}
H_T=\left(
\begin{array}{cc}
0 & V\\
V & 0
\end{array}
\right),
\end{equation}
allows tunneling through the barrier. Here we have chosen the 
relative
phases of the
basis states
$|S\rangle$ and $|N\rangle$ such that $V$ is positive, without loss of generality. Together, $H_W$ and $H_T$ form a simple problem common to many introductory quantum mechanics texts.

The remaining term, $H_D=H_{EM}+H_c$, gives the electromagnetic decay of the
two levels. $H_{EM}$ is composed of the electromagnetic (harmonic oscillator)
modes of the environment, and $H_c$ gives their couplings to the nucleus. In
this case, it is not necessary to treat the particulars of these terms; rather we work at the level of the experimentally determined decay rates $\Gamma_S/\hbar$ and $\Gamma_N/\hbar$~\cite{phenomnote}. The self-energy due to $H_D$ is
\begin{equation}
\Sigma=-\frac{i}{2}\left(
\begin{array}{cc}
\Gamma_S & 0\\
0 & \Gamma_N
\end{array}
\right).
\end{equation}

Dyson's Equation
\begin{equation}
G(E)=\left(\left[G_W(E)\right]^{-1}-H_T-\Sigma\right)^{-1}
\end{equation}
gives the Green's function of the full system, summing the effects of $\Sigma$ and $H_T$ to all orders. Treating the physics of the two wells ($G_W$), electromagnetic decay ($\Sigma$), and the barrier ($H_T$) on the same footing in this way is essential to a complete description of SD decay-out: all three play equally important roles in determining experimental observables, such as branching ratios.
The full Green's function of the two-level model is thus
\begin{eqnarray}
G & \equiv & \left(
\begin{array}{cc}
G_{SS} & G_{SN}\\
G_{NS} & G_{NN}
\end{array}
\right)\nonumber\\
& = & \left[(E-\varepsilon_S+i\Gamma_S/2)(E-\varepsilon_N+i\Gamma_N/2)-V^2\right]^{-1}\nonumber\\
& & \times\left(
\begin{array}{cc}
E-\varepsilon_N+i\Gamma_N/2 & V\\
V & E-\varepsilon_S+i\Gamma_S/2
\end{array}
\right).
\end{eqnarray}

At $t=0$, the nucleus is localized in the SD well by virtue of its previous,
measurable $E2$ decay. For later times, then, the probability to find the
nucleus in the SD or ND well
is given by $P_{i}(t)=|\tilde{G}_{iS}(t)|^2$, where $i=S,N$ respectively. Here
\begin{equation}
\tilde{G}_{iS}(t)=\int^\infty_{-\infty}\frac{dE}{2\pi}G_{iS}(E)\mathrm{e}^{-iEt/\hbar},
\end{equation}
the Fourier transform of $G_{ij}(E)$, is the retarded propagator from $S$ to $i$.

The resulting probabilities are
\begin{equation}
\label{PN}
P_N(t)=\frac{2V^2}{|\hbar\omega|^2}\mathrm{e}^{-\overline{\Gamma}t/\hbar}\left[\cosh\left(\omega_it\right)-\cos\left(\omega_rt\right)\right]
\end{equation}
and
\begin{equation}
\label{PS}
\begin{split}
P_S(t)=\frac{V^2}{|\hbar\omega|^2}\mathrm{e}^{-\overline{\Gamma}t/\hbar}\left(\frac{\hbar\omega_i+\Gamma'}{\Gamma'-\hbar\omega_i}\mathrm{e}^{\omega_it}+\frac{\Gamma'-\hbar\omega_i}{\hbar\omega_i+\Gamma'}\mathrm{e}^{-\omega_it}+\right.\\
\left.+\frac{i\hbar\omega_r+\Gamma'}{i\hbar\omega_r-\Gamma'}\mathrm{e}^{i\omega_rt}+\frac{i\hbar\omega_r-\Gamma'}{i\hbar\omega_r+\Gamma'}\mathrm{e}^{-i\omega_rt}\right).
\end{split}
\end{equation}
Here, $\overline{\Gamma}\equiv(\Gamma_N+\Gamma_S)/2$ and
$\Gamma'\equiv(\Gamma_N-\Gamma_S)/2$, while the real and imaginary parts of
the complex Rabi frequency
$\omega$ are given by 
\begin{equation}
\omega_{r,i}^2=\frac{\sqrt{\Omega^2+4\Delta^2\Gamma'^2}\pm\Omega}{2\hbar^2},\quad\Omega\equiv 4V^2+\Delta^2-\Gamma'^2,
\end{equation}
respectively, where the ``$+$'' sign is used for the real part, and the ``$-$'' for the imaginary.
As Eqs.~(\ref{PN}) and (\ref{PS}) demonstrate, $\omega_i$ is associated with decoherence due to coupling with the electromagnetic field, while $\omega_r$ is analogous to the real Rabi frequency of a closed two-well system.

The branching ratios 
$F_i=\left(\Gamma_i/\hbar\right)\int_0^\infty dtP_i(t)$ are found by time-integrating the
probabilities. The results are \cite{SB}
\begin{subequations}
\label{branchingratios}

\begin{equation}
\label{FN}
F_N  =  \frac{\Gamma_N\Gamma^\downarrow/\left(\Gamma_N+\Gamma^\downarrow\right)}
{\Gamma_S+\Gamma_N\Gamma^\downarrow/\left(\Gamma_N+\Gamma^\downarrow\right)}
\end{equation}
\begin{equation}
\label{FS}
F_S  =  \frac{\Gamma_S}
{\Gamma_S+\Gamma_N\Gamma^\downarrow/\left(\Gamma_N+\Gamma^\downarrow\right)},
\end{equation}

\end{subequations}
where
\begin{equation}
\label{Gammadowndef}
\Gamma^\downarrow\equiv\frac{2\overline{\Gamma}V^2}{\Delta^2+\overline{\Gamma}^2}.
\end{equation}
Equations (\ref{branchingratios}) are the expected results for series decay
out of a two-level problem. In this light, it is clear that
$\Gamma^\downarrow$ is simply the net rate for the nucleus, starting in the SD
well, to tunnel irreversibly through the barrier.

These results allow us to extract information about the potential barrier from
experiment. In particular, the values determined by a typical SD decay
experiment are $F_S=1-F_N$ and $\Gamma_S$, while $\Gamma_N$ can be estimated
by applying the cranking model to a Fermi gas density of states \cite{dossing95a}. From
Eq.~(\ref{branchingratios}), we find
\begin{equation}
\label{Gammadown}
\Gamma^\downarrow=
\Gamma_S\bigg/\left(\frac{F_S}{F_N}-\frac{\Gamma_S}{\Gamma_N}\right).
\end{equation}

\subsection{Determination of $V$}
\label{extractV}
To uniquely extract $V$ itself requires $\Delta$, which in turn implies
detailed knowledge of the spectrum in the ND well. In the
absence of this, 
we consider the entire statistical ensemble of two-level models, each
characterized 
by a different value of the unknown variable $\Delta$. 

To proceed, we construct a probability density function $\mathcal{P}(\Delta)$,
which gives the statistical weight each value of $\Delta$ has in the
ensemble.
The simplest ansatz for the distribution of energy levels in the ND well is the
``structureless'' Wigner surmise \cite{mehta}:
\begin{equation}
\label{Wignersurmise}
P(s)=\frac{\pi}{2}se^{-\pi s^2/4},
\end{equation}
where $s$ is the level spacing in units of its average value $D_N$. 
$\Delta$ is the detuning between the SD state and its nearest ND neighbor;
thus its magnitude must be less than half the spacing $sD_N$ between
the ND levels just above and just below the SD level.
Given this spacing, therefore, $\Delta$ is drawn from the
rectangular probability density
\begin{equation}
\mathcal{P}_s(\Delta)=\frac{1}{sD_N}\Theta\left(\frac{s}{2}-\frac{|\Delta|}{D_N}\right),
\end{equation}
where $\Theta$ is the Heaviside step function. The total probability
theorem yields the desired result \cite{CSBPRL}:
\begin{equation}
\label{GOEDelta}
\mathcal{P}(\Delta)=\int_0^\infty\mathcal{P}_s(\Delta)P(s)ds=\frac{\pi}{2D_N}\mathrm{erfc}\left(\sqrt{\pi}\frac{|\Delta|}{D_N}\right),
\end{equation}
where $\mathrm{erfc}(x)$ denotes the complementary error function of $x$.

A probabilistic statement like Eq.~(\ref{GOEDelta}) obviates the need for
exact knowledge of $\Delta$. 
To arrive at the probability density function $\mathcal{P}(V)$,
one need only perform an elementary change of variables:
\begin{equation}
\mathcal{P}(V)=2\mathcal{P}(\Delta)\left|\frac{d\Delta(V)}{dV}\right|,
\end{equation}
where the factor 2 results from our choice of phase for $V$.
Here, $\left|\Delta(V)\right|$ is a function, not the random
variable $\Delta$; it is found from Eq.~(\ref{Gammadowndef}) to be
\begin{equation}
\label{DeltaofV}
\left|\Delta(V)\right|=\sqrt{\frac{2\overline{\Gamma}}{\Gamma^\downarrow}\left(V^2-V_{min}^2\right)},
\end{equation}
where $V_{min}=\sqrt{\frac{1}{2}\Gamma^\downarrow\overline{\Gamma}}$ is the
smallest $V$ consistent with the two-level model. $\mathcal{P}(V)$ is thus
seen to be \cite{CSBPRL,deltanote}
\begin{equation}
\label{PofV}
\mathcal{P}(V)=\left\{\begin{array}{ll}
\frac{2\pi}{D_N}\frac{\overline{\Gamma}V}{\Gamma^\downarrow|\Delta(V)|}\mathrm{erfc}\left(\sqrt{\pi}\frac{|\Delta(V)|}{D_N}\right), & V> V_{min}\\
0 & \textrm{otherwise}
\end{array}\right. .
\end{equation}

A probability distribution such as Eq.~(\ref{PofV}) represents the most one can say
about $V$ without microscopic knowledge of the ND well. 
The mean of $\mathcal{P}(V)$
is
\begin{equation}
\langle
V\rangle=\sqrt{\frac{\Gamma^\downarrow}{2\overline{\Gamma}}}\left[\frac{D_N}{4}+\mathcal{O}\left(\frac{\overline{\Gamma}^2}{D_N}\right)\right],
\end{equation}
while the standard deviation is
\begin{equation}
\sigma_V=\sqrt{\frac{\Gamma^\downarrow}{2\overline{\Gamma}}}\left[D_N\sqrt{\frac{1}{3\pi}-\frac{1}{16}}+\mathcal{O}\left(\overline{\Gamma}\right)\right].
\end{equation}
$\sigma_V/\langle V\rangle\approx 84\%$, indicating that $\mathcal{P}(V)$ is
well peaked about $\langle V\rangle$, and thus this mean provides a good
measure of the likely value of $V$.

The Wigner surmise (\ref{Wignersurmise}), provides a reasonable, neutral guess
at the spacings of states in the ND well. It is closely related, and
may be considered a good approximation, to the level distribution of the
Gaussian orthogonal ensemble \cite{mehta}. Nevertheless, it is straightforward
to reproduce the preceeding analysis, substituting a level-spacing
density of choice for Eq.~(\ref{Wignersurmise}).

\section{Results for the 150 and 190 Mass Regions}
\label{results}
Table \ref{bigtable} gives the values of $\Gamma^\downarrow$ and $\langle
V\rangle$ for all SD decays for which the four parameters, $F_N$; $\Gamma_S$;
$\Gamma_N$; and $D_N$, are known. In the table, we have further defined the
series rate to irreversibly leave the SD band: 
\begin{equation}
\Gamma_{out}/\hbar=\frac{\Gamma_N\Gamma^\downarrow}{\Gamma_N+\Gamma^\downarrow}/\hbar=\Gamma_S\frac{F_N}{F_S}/\hbar.
\end{equation}
It is $\Gamma_{out}$, directly extractable from experimental results, 
which competes with $\Gamma_S$ to determine whether a nucleus
will decay out of or remain within the SD band. 

\begin{table*}
\caption{Results of the two-level model, for all SD decays for which
  sufficient data (branching ratios, $\Gamma_S$, $\Gamma_N$, and $D_N$) are
  known. $I$ is the nuclear spin quantum number. Note that $\langle
  V_c\rangle$ and $\Gamma_S/\left(\Gamma_S+\Gamma_N\right)$ do not depend on
  $F_S$. The rightmost column gives the sources of the experimental inputs and
  the estimates of $\Gamma_N$ and $D_N$.}
\label{bigtable}
\begin{tabular}{|c||r@{.}lr@{.}lr@{.}lr@{.}l|r@{.}lr@{.}lr@{.}l|r@{.}lr@{.}l@{\hspace{1em}}|c|}
\hline\hline
nucleus($I$) & \multicolumn{2}{c}{$F_S$} & \multicolumn{2}{c}{$\Gamma_S$} &
\multicolumn{2}{c}{$\Gamma_N$} & \multicolumn{2}{c}{$D_N$} &
\multicolumn{2}{c}{$\Gamma^\downarrow$} & \multicolumn{2}{c}{$\Gamma_{out}$} &
\multicolumn{2}{c}{$\langle V\rangle$} & \multicolumn{2}{c}{$\langle
  V_c\rangle$} & 
\multicolumn{2}{c}{$\frac{\Gamma_S}{\Gamma_S+\Gamma_N}$} & Refs.\\
& \multicolumn{2}{c}{} & \multicolumn{2}{c}{(meV)} & \multicolumn{2}{c}{(meV)}
& \multicolumn{2}{c}{(eV)} & \multicolumn{2}{c}{(meV)} &
\multicolumn{2}{c}{(meV)} &
\multicolumn{2}{c}{(eV)} & \multicolumn{2}{c}{(eV)} & \multicolumn{2}{c}{} & \\
\hline
\hline
${}^{192}$Hg(12) & 0&74  & 0&128 & 0&613 & 135& & 0&049 & 0&045 & 8&7 & 14&0 &
0&173 & \cite{lauritsen00a,wilson05a}\\ 
${}^{192}$Hg(10) & 0&08  & 0&050 & 0&733 & 89& & 2&7 & 0&58 & 41& & 5&6 &
0&064 & \cite{lauritsen00a,wilson05a}\\ 
\hline
${}^{192}$Pb(16) & $>$0&99 & 0&487 & 0&192 & 1,362& & $<$0&0050 & $<$0&0049 &
$<$29& & 288& & 0&717 & \cite{wilson03a,wilson04a}\\ 
${}^{192}$Pb(14) & 0&98 & 0&266 & 0&201 & 1,258& & 0&0056 & 0&0054 & 34& &
237& &
0&570 & \cite{wilson03a,wilson04a}\\ 
${}^{192}$Pb(12) & 0&66 & 0&132 & 0&200 & 1,272& & 0&10 & 0&067 & 170& & 201& 
& 0&398 & \cite{wilson03a,wilson04a}\\ 
${}^{192}$Pb(10) & 0&12 & 0&048 & 0&188 & 1,410& & 1&9$\dagger$ & 0&35 &
1000&$\dagger$ & 160& 
& 0&203 & \cite{wilson03a,wilson04a}\\ 
${}^{192}$Pb(8) & $<$0&25 & 0&016 & 0&169 & 1,681& & $>$0&067 & $>$0&048 &
$>$250& & 120& &
0&086 & \cite{wilson03a,wilson04a}\\ 
\hline

${}^{194}$Hg(12) & 0&58 & 0&097 & 4&8 & 16&3 & 0&071 & 0&070 & 0&49 & 0&58 &
0&020 & \cite{khoo96a,kuehn97a,moore97a,kruecken01a}\\ 
${}^{194}$Hg(10) & $<$0&09 & 0&039 & 4&1 & 26&2 & $>$0&44 & $>$0&40 & $>$2&1 &
0&64 & 0&0094 & \cite{khoo96a,kuehn97a,moore97a,kruecken01a}\\ 
\hline
${}^{194}$Hg(12) & 0&60 & 0&108 & 21& & 344& & 0&072 & 0&072 & 5&0 & 6&1 &
0&0051 & \cite{lauritsen02a}\\ 
${}^{194}$Hg(10) & 0&03 & 0&046 & 20& & 493& & 1&6 & 1&5 & 35& & 5&9 & 0&0023 & \cite{lauritsen02a}\\ 
\hline
${}^{194}$Hg(12) & 0&60 & 0&086 & 1&345 & 19& & 0&060 & 0&057 & 0&97 & 1&2 & 
0&060 & \cite{moore97a,wilson05a}\\ 
${}^{194}$Hg(10) & $\le$0&05 & 0&033 & 1&487 & 14& & $\ge$1&1 & $\ge$0&63 &
$\ge$3&0 & 0&52 & 
0&022 & \cite{moore97a,wilson05a}\\ 
\hline
${}^{194}$Hg(15) & 0&90 & 0&230 & 4&0 & 26&5 & 0&026 & 0&026 & 0&52 & 1&5 &
0&054 & \cite{moore97a,kruecken01a}\\ 
${}^{194}$Hg(13) & 0&84 & 0&110 & 4&5 & 19&9 & 0&021 & 0&021 & 0&34 & 0&77 &
0&024 & \cite{moore97a,kruecken01a}\\ 
${}^{194}$Hg(11) & $<$0&07 & 0&048 & 6&4 & 7&2 & $>$0&71 & $>$0&64 & $>$0&60 &
0&15 & 0&0074 &
 \cite{moore97a,kruecken01a}\\ 
\hline
${}^{194}$Pb(10) & 0&90 & 0&045 & 0&08 & 21,700& & 0&0053 & 0&0050 & 1100& &
3300& & 0&36 & \cite{willsau93a,lopezmartens96a,hauschild97a,kruecken01a}\\ 
${}^{194}$Pb(8) & 0&62 & 0&014 & 0&50 & 2,200& & 0&0087 & 0&0086 & 72& & 90&
& 0&027 & \cite{willsau93a,lopezmartens96a,hauschild97a,kruecken01a}\\ 
${}^{194}$Pb(6) & $<$0&09 & 0&003 & 0&65 & 1,400& & $>$0&032 & $>$0&030 &
$>$77& & 20& & 
0&005 & \cite{willsau93a,lopezmartens96a,hauschild97a,kruecken01a}\\

\hline
${}^{194}$Pb(12) & $>$0&99 & 0&125 & 0&476 & 236& & $<$0&0013 & $<$0&0013 &
$<$2&7 & 26&9 & 
0&208 & \cite{kruecken01a,wilson04a}\\ 
${}^{194}$Pb(10) & 0&90 & 0&045 & 0&470 & 244& & 0&0051 & 0&0050 & 6&1 & 18& &
0&087 & \cite{kruecken01a,wilson04a}\\ 
${}^{194}$Pb(8) & 0&65 & 0&014 & 0&445 & 273& & 0&0077 & 0&0076 & 8&8 & 12& & 
0&031 & \cite{kruecken01a,wilson04a}\\ 
${}^{194}$Pb(6) & $<$0&04 & 0&003 & 0&405 & 333& & $>$0&088 & $>$0&072 & $>$39&
& 7& & 0&007 & \cite{kruecken01a,wilson04a}\\ 
\hline\hline
${}^{152}$Dy(28) & 0&60 & 10&0 & 17& & 220& & 11& & 6&7 & 35& & 33& & 0&37 & \cite{lauritsen02a}\\ 
${}^{152}$Dy(26) & 0&19 & 7&0 & 17& & 194& & 140&$\dagger$ & 30& & 120&$\dagger$
& 26& & 0&29 &
\cite{lauritsen02a}\\ 
\hline\hline
\end{tabular}\\
$\dagger${\footnotesize Calculated statistically, as explained in the appendix.}
\end{table*}

The dynamics of 
SD decay is a consequence of the strong separation of energy scales,
\begin{equation}
\Gamma_S, \Gamma_N \ll D_N,\langle V\rangle,
\end{equation}
exhibited in Table \ref{bigtable}.
In the 190 mass region, particularly, we find that
parameters relating to the potential double-well, such as $D_N$ and $V$, are
10's to 1000's of electron--Volts, while those relating to electromagnetic
decay are fractions of meV. This is to be expected, since nuclear forces are,
of course, many orders of magnitude stronger than electromagnetic ones.

Each of these energy scales
has a typical rate associated with it: oscillations within the
two-level system are characterized by $\omega_r$,
whereas $\overline{\Gamma}/\hbar$ gives the typical rate for electromagnetic
decay in Eqs.~(\ref{PN})--(\ref{PS}). Since $\hbar\omega_r\gg\overline{\Gamma}$, it is clear that SD decay is
primarily a coherent process: \emph{a nucleus generally undergoes thousands of
  virtual Rabi oscillations during a single decay event}. Only if both $V$ and
$\Delta$ were of order meV 
or smaller
could the decay be incoherent, and such
``accidental''
near-degeneracies are masked by the fact that, since $\Gamma^\downarrow\approx 0$,
when $V$ is very small, the nucleus cannot leave the SD band.
Moreover, the probability for $\Delta$ to be of order $\overline{\Gamma}$
is seen from Eq.~(\ref{GOEDelta}) to be $\sim\overline{\Gamma}/D_N$.

Equation (\ref{FS}) can be rewritten:
\begin{equation}
\label{FSintermsofratios}
F_S=1-\frac{1}{1+\left(V_c/V\right)^2+\Gamma_S/\Gamma_N},
\end{equation}
where
\begin{equation}
\label{Vc}
V_c^2\equiv\left(\Delta^2+\overline{\Gamma}^2\right)\frac{\Gamma_S/\Gamma_N}{1+\Gamma_S/\Gamma_N}.
\end{equation}
The critical tunneling matrix element $V_c$ can be
extracted from experiment via a statistical approach similar to that described
for $V$ in Section \ref{extractV}, except that the result does not depend on
$\Gamma^\downarrow$ or $F_S$. The resulting average values $\langle
V_c\rangle$ are tabulated in Table \ref{bigtable}.

According to Eq.~(\ref{FSintermsofratios}), only two dimensionless
parameters, $\Gamma_S/\Gamma_N$ and $V_c/V$, play a role in
determining the branching ratios; each corresponds to one of the problem's
two energy scales. It is clear from Table \ref{bigtable} that
the first, $\Gamma_S/\Gamma_N$, decreases dramatically over the course of
each SD band's decay-out. This can be understood physically as a relaxation of
the nucleus's centrifugal stretching, and consequent reduced coupling to the
electric quadrupole field, as its spin lowers.

We extracted $\langle V\rangle$ using the experimental branching ratios;
thus it would be circular to make use of those values in our discussion of
$F_S$'s universality. Instead we note that Eqs.~(\ref{FS}) and
(\ref{FSintermsofratios}) have the limit
\begin{equation}
\lim_{V\rightarrow\infty}F_S=\frac{\Gamma_S}{\Gamma_S+\Gamma_N},
\end{equation}
and that, in the two-level model, $F_S$ is a monotonically decreasing function
of $V$. Values of this limit are given in Table \ref{bigtable}. As we move
down each decay chain, it is clear that the
experimental branching ratios converge to these values, and hence we conclude
that $V_c/V$, too, is decreasing quickly during each decay chain. This
also is to be expected: $V$ depends exponentially on the barrier
height \cite{kruecken96a}, which decreases as the ND well drops further below the SD one in
energy.

Figure \ref{fig:results} shows $F_S$ as a function of $V_c/V$
and $\Gamma_S/\Gamma_N$. 
As both $\Gamma_S/\Gamma_N$ and $V_c/V$ decrease, the SD branching ratio
decreases to an abrupt plateau.
Furthermore, within the decay-allowed region
\begin{equation}
\Gamma_N\gtrsim\Gamma_S\,\cap\,V\gtrsim V_c
\end{equation}
all the curves quickly become nearly identical.

Thus, we explain the universal nature of the
decay-out profiles as follows: SD decay-out is only allowed when 
suitably low values of both $\Gamma_S/\Gamma_N$ and $V_c/V$ are achieved,
both of which decrease quickly with decreasing spin. The nucleus, therefore,
enters the region of allowed decay-out very suddenly, moving down the
curves from a case of very high in-band intensity to one of almost complete
decay-out;
high in-band intensity
corresponds to the long chains of pre-decay SD states observed
in experiment, 
while the 
rapid transition into the decay-allowed region forms the abrupt decay
profiles. However, once the system enters the decay-allowed region, the
branching ratios saturate, and hence are no longer sensitive to further changes in the parameters, as
Fig.~\ref{fig:results} shows. The decay profile is consequently universal.

If ${}^{152}\mathrm{Dy}$ is a representative example, the variation of
$\Gamma_S/\Gamma_N$ in the 150 mass region is somewhat slower than in the 190
region, although
$V_c/V$ still changes dramatically, as seen from $F_S$'s approach to its
$V\rightarrow\infty$ limit. Since rapid 
entry into the decay-allowed region is a necessary precondition of
universal behavior, we expect that, as more data become available for these
nuclei, a somewhat lesser degree of universality will be observed.

\section{Conclusions}
\label{conclusions}
The two-level model has elsewhere \cite{CSBPRL,dzyublik03a} been shown to be the simplest description of
the SD decay-out process which still encapsulates the essential physics. It
describes a two-step decay: first, the nucleus undergoes mainly
coherent Rabi oscillations between the SD and ND wells, after which it finally
decays into one band or the other. By using a statistical approach, the
two-level model can extract as much information as is
possible from decay experiments, including the Hamiltonian matrix element for
tunneling through the potential barrier, which is of direct relevance to
nuclear structure. Table \ref{bigtable} demonstrates the results of this technique
for all decays with sufficient data, to date.

Moreover, the most striking property of SD decay-out in the 190
mass region,
universality of the decay profiles, is seen to correspond to universal
behavior of the branching ratio in the two-level model.
The two-level branching ratio is completely determined by two dimensionless
parameters, $\Gamma_S/\Gamma_N$ and $V_c/V$, each of which corresponds to one
of the two disparate energy scales of the problem.
Decay-out only occurs when both of these 
parameters 
are decreasing rapidly.  Thus,
in the sector of parameter space for which decays are allowed, the branching
ratios saturate and are insensitive to variations of the parameters.
Assuming that the relationship between nuclear spin and barrier shape is
reasonably consistent from nucleus to nucleus, the resulting decay profiles are
necessarily similar.

\appendix
\section*{Appendix}
Equation (\ref{Gammadown}) places a limit on the
experimentally-determined quantities. Positivity of $\Gamma^\downarrow$
requires that
\begin{equation}
\Gamma_N>\Gamma_{out}
\end{equation}
In only two decays of Table \ref{bigtable}, ${}^{192}\mathrm{Pb}(10)$ and ${}^{152}\mathrm{Dy}(26)$, is this condition 
violated. While it is possible that this is due to a breakdown of the
two-level approximation in these cases, in the absence of a physical
argument for the near degeneracy of two or more ND levels, it is 
more
likely that one or more of the input parameters is poorly known. $\Gamma_N$, in
particular, is difficult to estimate, with uncertainty
$\sigma_{\Gamma_N}\sim\Gamma_N$. 

Thus, we estimate $\Gamma^\downarrow$ statistically for these two decays, assuming the true $\Gamma_N$ differs
from the estimated value $\Gamma_N^0$ by a ``cut'' normal distribution:
\begin{equation}
\mathcal{P}(\Gamma_N)=\left\{\begin{array}{ll}
\frac{\mathcal{A}}{\Gamma_N^0\sqrt{2\pi}}\mathrm{e}^{-\left(\frac{\Gamma_N-\Gamma_N^0}{\sqrt{2}\Gamma_N^0}\right)^2},
  & \Gamma_N>\Gamma_{out}\\
  0 & \textrm{otherwise}
\end{array}\right.,
\end{equation}
where the constant of renormalization due to the constraint is
\begin{equation}
\mathcal{A}=2\left\{\mathrm{erfc}\left[\frac{1}{\sqrt{2}}\left(\frac{\Gamma_{out}}{\Gamma_N^0}-1\right)\right]\right\}^{-1}.
\end{equation}
Assuming that the two-level approximation is valid, the probability density
function of $\Gamma^\downarrow$ follows:
\begin{equation}
\mathcal{P}(\Gamma^\downarrow)=\mathcal{P}(\Gamma_N)\left|\frac{d\Gamma_N}{d\Gamma^\downarrow}\right|=\left(\frac{\Gamma_N}{\Gamma^\downarrow}\right)^2\mathcal{P}(\Gamma_N),
\end{equation}
where $\Gamma^\downarrow$ is the function of $\Gamma_N$ given by
Eq.~(\ref{Gammadown}). For ${}^{192}\mathrm{Pb}(10)$ and
${}^{152}\mathrm{Dy}(26)$, Table~\ref{bigtable} gives the median of this
distribution as the typical value of $\Gamma^\downarrow$, from which
$\langle V\rangle$ is found.

\section*{Acknowledgments}
We thank Anna Wilson, Teng Lek Khoo, Daniel Stein, J\'er\^ome B\"urki, and
Bertrand Giraud for useful discussions, and TRIUMF for hospitality during the
formation of part of this manuscript. This work was partially funded by United
States NSF grants PHY-0244389 and PHY-0555396.

\bibliography{SDUNT}

\begin{thebibliography}{27}
\expandafter\ifx\csname natexlab\endcsname\relax\def\natexlab#1{#1}\fi
\expandafter\ifx\csname bibnamefont\endcsname\relax
  \def\bibnamefont#1{#1}\fi
\expandafter\ifx\csname bibfnamefont\endcsname\relax
  \def\bibfnamefont#1{#1}\fi
\expandafter\ifx\csname citenamefont\endcsname\relax
  \def\citenamefont#1{#1}\fi
\expandafter\ifx\csname url\endcsname\relax
  \def\url#1{\texttt{#1}}\fi
\expandafter\ifx\csname urlprefix\endcsname\relax\def\urlprefix{URL }\fi
\providecommand{\bibinfo}[2]{#2}
\providecommand{\eprint}[2][]{\url{#2}}

\bibitem{aberg90}
\bibinfo{author}{\bibfnamefont{S.}~\bibnamefont{{\AA}berg}},
  \bibinfo{journal}{Nucl.\ Phys.\ A} {\bibinfo{volume}{520}}
  (\bibinfo{year}{1990}) \bibinfo{pages}{35c}.

\bibitem{twin90}
\bibinfo{author}{\bibfnamefont{P.~J.} \bibnamefont{Twin}},
  \bibinfo{journal}{Nucl.\ Phys.\ A} {\bibinfo{volume}{520}}
  (\bibinfo{year}{1990}) \bibinfo{pages}{17c}.

\bibitem{nolan88}
\bibinfo{author}{\bibfnamefont{P.~J.} \bibnamefont{Nolan}} \bibnamefont{and}
  \bibinfo{author}{\bibfnamefont{P.~J.} \bibnamefont{Twin}},
  \bibinfo{journal}{Annu.\ Rev.\ Nucl.\ Part.\ Sci.}
  {\bibinfo{volume}{38}} (\bibinfo{year}{1988}) \bibinfo{pages}{533};
\bibinfo{author}{\bibfnamefont{R.~V.~F.} \bibnamefont{Janssens}}
  \bibnamefont{and} \bibinfo{author}{\bibfnamefont{T.~L.} \bibnamefont{Khoo}},
  \bibinfo{journal}{Annu.\ Rev.\ Nucl.\ Part.\ Sci.}
  {\bibinfo{volume}{41}} (\bibinfo{year}{1991}) \bibinfo{pages}{321};
\bibinfo{author}{\bibfnamefont{J.~F.} \bibnamefont{Sharpey-Schafer}},
  \bibinfo{journal}{Prog.\ Part.\ Nucl.\ Phys.} {\bibinfo{volume}{28}}
  (\bibinfo{year}{1992}) \bibinfo{pages}{187};
\bibinfo{author}{\bibfnamefont{T.~L.} \bibnamefont{Khoo}}, in
  \emph{\bibinfo{booktitle}{Tunneling in Complex Systems\emph{, Proc.\ from the
  INT, Vol.\ 5}}}, edited by
  \bibinfo{editor}{\bibfnamefont{S.}~\bibnamefont{Tomsovic}}
  (\bibinfo{publisher}{World Scientific}, \bibinfo{address}{Singapore},
  \bibinfo{year}{1998}), p. \bibinfo{pages}{229}.

\bibitem{wilson05a}
\bibinfo{author}{\bibfnamefont{A.~N.} \bibnamefont{Wilson}},
  \bibinfo{author}{\bibfnamefont{A.~J.} \bibnamefont{Sargeant}},
  \bibinfo{author}{\bibfnamefont{P.~M.} \bibnamefont{Davidson}},
  \bibnamefont{and} \bibinfo{author}{\bibfnamefont{M.~S.}
  \bibnamefont{Hussein}}, \bibinfo{journal}{Phys.\ Rev.\ C}
  {\bibinfo{volume}{71}} (\bibinfo{year}{2005}) \bibinfo{pages}{034319}.

\bibitem{shimizu93}
Y. R. Schimizu, E. Vigezzi, T. D{\o}ssing, and R. A. Broglia, Nucl.\
  Phys.\ A 557 (1993) 99c.

\bibitem{vigezzi90a}
\bibinfo{author}{\bibfnamefont{E.}~\bibnamefont{Vigezzi}},
  \bibinfo{author}{\bibfnamefont{R.~A.} \bibnamefont{Broglia}},
  \bibnamefont{and}
  \bibinfo{author}{\bibfnamefont{T.}~\bibnamefont{D{\o}ssing}},
  \bibinfo{journal}{Nucl.\ Phys.\ A} {\bibinfo{volume}{520}}
  (\bibinfo{year}{1990}{\natexlab{a}}) \bibinfo{pages}{179c};
\bibinfo{author}{\bibfnamefont{E.}~\bibnamefont{Vigezzi}},
  \bibinfo{author}{\bibfnamefont{R.~A.} \bibnamefont{Broglia}},
  \bibnamefont{and}
  \bibinfo{author}{\bibfnamefont{T.}~\bibnamefont{D{\o}ssing}},
  \bibinfo{journal}{Phys.\ Lett.\ B} {\bibinfo{volume}{249}}
  (\bibinfo{year}{1990}{\natexlab{b}}) \bibinfo{pages}{163}.

\bibitem{khoo93a}
\bibinfo{author}{\bibfnamefont{T.~L.} \bibnamefont{Khoo}},
  \bibinfo{author}{\bibfnamefont{T.}~\bibnamefont{Lauritsen}},
  \bibinfo{author}{\bibfnamefont{I.}~\bibnamefont{Ahmad}},
  \bibinfo{author}{\bibfnamefont{M.~P.} \bibnamefont{Carpenter}},
  \bibinfo{author}{\bibfnamefont{P.~B.} \bibnamefont{Fernandez}},
  \bibinfo{author}{\bibfnamefont{R.~V.~F.} \bibnamefont{Janssens}},
  \bibinfo{author}{\bibfnamefont{E.~F.} \bibnamefont{Moore}},
  \bibinfo{author}{\bibfnamefont{F.~L.~H.} \bibnamefont{Wolfs}},
  \bibinfo{author}{\bibnamefont{{Ph.\ Benet}}},
  \bibinfo{author}{\bibfnamefont{P.~J.} \bibnamefont{Daly}},
  \bibnamefont{\emph{et~al.}}, \bibinfo{journal}{Nucl.\ Phys.\ A}
  {\bibinfo{volume}{557}} (\bibinfo{year}{1993}) \bibinfo{pages}{83c}.

\bibitem{SB}
\bibinfo{author}{\bibfnamefont{C.~A.} \bibnamefont{Stafford}} \bibnamefont{and}
  \bibinfo{author}{\bibfnamefont{B.~R.} \bibnamefont{Barrett}},
  \bibinfo{journal}{Phys.\ Rev.\ C} {\bibinfo{volume}{60}}
  (\bibinfo{year}{1999}) \bibinfo{pages}{051305(R)}.

\bibitem{CSBPRL}
\bibinfo{author}{\bibfnamefont{D.~M.} \bibnamefont{Cardamone}},
  \bibinfo{author}{\bibfnamefont{C.~A.} \bibnamefont{Stafford}},
  \bibnamefont{and} \bibinfo{author}{\bibfnamefont{B.~R.}
  \bibnamefont{Barrett}}, \bibinfo{journal}{Phys.\ Rev.\ Lett.}
  {\bibinfo{volume}{91}} (\bibinfo{year}{2003}) \bibinfo{pages}{102502}.

\bibitem{dzyublik03a}
\bibinfo{author}{\bibnamefont{{A. Ya.\ Dzyublik}}} \bibnamefont{and}
  \bibinfo{author}{\bibfnamefont{V.~V.} \bibnamefont{Utyuzh}},
  \bibinfo{journal}{Phys.\ Rev.\ C} {\bibinfo{volume}{68}}
  (\bibinfo{year}{2003}) \bibinfo{pages}{024311}.

\bibitem{phenomnote}
The central role of electromagnetic processes in SD nuclei
  decay suggests such a
phenomenological approach. A microscopic theory, while doubtless
desirable in its own right, would also obscure much of the
experimentally-relevant dynamics.

\bibitem{dossing95a}
\bibinfo{author}{\bibfnamefont{T.}~\bibnamefont{D{\o}ssing}} \bibnamefont{and}
  \bibinfo{author}{\bibfnamefont{E.}~\bibnamefont{Vigezzi}},
  \bibinfo{journal}{Nucl.\ Phys.\ A} {\bibinfo{volume}{587}}
  (\bibinfo{year}{1995}) \bibinfo{pages}{13}.

\bibitem{mehta}
\bibinfo{author}{\bibfnamefont{M.~L.} \bibnamefont{Mehta}},
  \emph{\bibinfo{title}{Random Matrices and the Statistical Theory of Energy
  Levels}} (\bibinfo{publisher}{Academic}, \bibinfo{address}{New York},
  \bibinfo{year}{1967}).

\bibitem{deltanote}
Note that $\mathcal{P}(V)$ explicitly does \emph{not} assume any
particualr value of $\Delta$, as was claimed in Ref.~\cite{wilson05a}. Rather,
it represents a weighted average over all values of $\Delta$.

\bibitem{lauritsen00a}
\bibinfo{author}{\bibfnamefont{T.}~\bibnamefont{Lauritsen}},
  \bibinfo{author}{\bibfnamefont{T.~L.} \bibnamefont{Khoo}},
  \bibinfo{author}{\bibfnamefont{I.}~\bibnamefont{Ahmad}},
  \bibinfo{author}{\bibfnamefont{M.~P.} \bibnamefont{Carpenter}},
  \bibinfo{author}{\bibfnamefont{R.~V.~F.} \bibnamefont{Janssens}},
  \bibinfo{author}{\bibfnamefont{A.}~\bibnamefont{Korichi}},
  \bibinfo{author}{\bibfnamefont{A.}~\bibnamefont{Lopez-Martens}},
  \bibinfo{author}{\bibfnamefont{H.}~\bibnamefont{Amro}},
  \bibinfo{author}{\bibfnamefont{S.}~\bibnamefont{Berger}},
  \bibinfo{author}{\bibfnamefont{L.}~\bibnamefont{Calderin}},
  \bibnamefont{\emph{et~al.}}, \bibinfo{journal}{Phys.\ Rev.\ C}
  {\bibinfo{volume}{62}} (\bibinfo{year}{2003}) \bibinfo{pages}{044316}.

\bibitem{wilson03a}
\bibinfo{author}{\bibfnamefont{A.~N.} \bibnamefont{Wilson}},
  \bibinfo{author}{\bibfnamefont{G.~D.} \bibnamefont{Dracoulis}},
  \bibinfo{author}{\bibfnamefont{A.~P.} \bibnamefont{Byrne}},
  \bibinfo{author}{\bibfnamefont{P.~M.} \bibnamefont{Davidson}},
  \bibinfo{author}{\bibfnamefont{G.~J.} \bibnamefont{Lane}},
  \bibinfo{author}{\bibfnamefont{R.~M.} \bibnamefont{Clark}},
  \bibinfo{author}{\bibfnamefont{P.}~\bibnamefont{Fallon}},
  \bibinfo{author}{\bibfnamefont{A.}~\bibnamefont{G{\"o}rgen}},
  \bibinfo{author}{\bibfnamefont{A.~O.} \bibnamefont{Macchiavelli}},
  \bibnamefont{and} \bibinfo{author}{\bibfnamefont{D.}~\bibnamefont{Ward}},
  \bibinfo{journal}{Phys.\ Rev.\ Lett.} {\bibinfo{volume}{90}}
  (\bibinfo{year}{2003}) \bibinfo{pages}{142501}.

\bibitem{wilson04a}
\bibinfo{author}{\bibfnamefont{A.~N.} \bibnamefont{Wilson}} \bibnamefont{and}
  \bibinfo{author}{\bibfnamefont{P.~M.} \bibnamefont{Davidson}},
  \bibinfo{journal}{Phys.\ Rev.\ C} {\bibinfo{volume}{69}}
  (\bibinfo{year}{2004}) \bibinfo{pages}{041303(R)}.

\bibitem{khoo96a}
\bibinfo{author}{\bibfnamefont{T.~L.} \bibnamefont{Khoo}},
  \bibinfo{author}{\bibfnamefont{M.~P.} \bibnamefont{Carpenter}},
  \bibinfo{author}{\bibfnamefont{T.}~\bibnamefont{Lauritsen}},
  \bibinfo{author}{\bibfnamefont{D.}~\bibnamefont{Ackermann}},
  \bibinfo{author}{\bibfnamefont{I.}~\bibnamefont{Ahmad}},
  \bibinfo{author}{\bibfnamefont{D.~J.} \bibnamefont{Blumenthal}},
  \bibinfo{author}{\bibfnamefont{S.~M.} \bibnamefont{Fischer}},
  \bibinfo{author}{\bibfnamefont{R.~V.~F.} \bibnamefont{Janssens}},
  \bibinfo{author}{\bibfnamefont{D.}~\bibnamefont{Nisius}},
  \bibinfo{author}{\bibfnamefont{E.~F.} \bibnamefont{Moore}},
  \bibnamefont{\emph{et~al.}}, \bibinfo{journal}{Phys.\ Rev.\ Lett.}
  {\bibinfo{volume}{76}} (\bibinfo{year}{1996}) \bibinfo{pages}{1583}.

\bibitem{kuehn97a}
\bibinfo{author}{\bibfnamefont{R.}~\bibnamefont{K{\"u}hn}},
  \bibinfo{author}{\bibfnamefont{A.}~\bibnamefont{Dewald}},
  \bibinfo{author}{\bibfnamefont{R.}~\bibnamefont{Kr{\"u}cken}},
  \bibinfo{author}{\bibfnamefont{C.}~\bibnamefont{Meier}},
  \bibinfo{author}{\bibfnamefont{R.}~\bibnamefont{Peusquens}},
  \bibinfo{author}{\bibfnamefont{H.}~\bibnamefont{Tiessler}},
  \bibinfo{author}{\bibfnamefont{O.}~\bibnamefont{Vogel}},
  \bibinfo{author}{\bibfnamefont{S.}~\bibnamefont{Kasemann}},
  \bibinfo{author}{\bibfnamefont{P.}~\bibnamefont{von Brentano}},
  \bibinfo{author}{\bibfnamefont{D.}~\bibnamefont{Bazzacco}},
  \bibnamefont{\emph{et~al.}}, \bibinfo{journal}{Phys.\ Rev.\ C}
  {\bibinfo{volume}{55}} (\bibinfo{year}{1997}) \bibinfo{pages}{R1002}.

\bibitem{moore97a}
\bibinfo{author}{\bibfnamefont{E.~F.} \bibnamefont{Moore}},
  \bibinfo{author}{\bibfnamefont{T.}~\bibnamefont{Lauritsen}},
  \bibinfo{author}{\bibfnamefont{R.~V.~F.} \bibnamefont{Janssens}},
  \bibinfo{author}{\bibfnamefont{T.~L.} \bibnamefont{Khoo}},
  \bibinfo{author}{\bibfnamefont{D.}~\bibnamefont{Ackermann}},
  \bibinfo{author}{\bibfnamefont{I.}~\bibnamefont{Ahmad}},
  \bibinfo{author}{\bibfnamefont{H.}~\bibnamefont{Amro}},
  \bibinfo{author}{\bibfnamefont{D.}~\bibnamefont{Blumenthal}},
  \bibinfo{author}{\bibfnamefont{M.~P.} \bibnamefont{Carpenter}},
  \bibinfo{author}{\bibfnamefont{S.~M.} \bibnamefont{Fischer}},
  \bibnamefont{\emph{et~al.}}, \bibinfo{journal}{Phys.\ Rev.\ C}
  {\bibinfo{volume}{55}} (\bibinfo{year}{1997}) \bibinfo{pages}{R2150}.

\bibitem{kruecken01a}
\bibinfo{author}{\bibfnamefont{R.}~\bibnamefont{Kr{\"u}cken}},
  \bibinfo{author}{\bibfnamefont{A.}~\bibnamefont{Dewald}},
  \bibinfo{author}{\bibfnamefont{P.}~\bibnamefont{von Brentano}},
  \bibnamefont{and} \bibinfo{author}{\bibfnamefont{H.~A.}
  \bibnamefont{Weidenm{\"u}ller}}, \bibinfo{journal}{Phys.\ Rev.\ C}
  {\bibinfo{volume}{64}} (\bibinfo{year}{2001}) \bibinfo{pages}{064316}.

\bibitem{lauritsen02a}
\bibinfo{author}{\bibfnamefont{T.}~\bibnamefont{Lauritsen}},
  \bibinfo{author}{\bibfnamefont{M.~P.} \bibnamefont{Carpenter}},
  \bibinfo{author}{\bibfnamefont{T.}~\bibnamefont{D{\o}ssing}},
  \bibinfo{author}{\bibfnamefont{P.}~\bibnamefont{Fallon}},
  \bibinfo{author}{\bibfnamefont{B.}~\bibnamefont{Herskind}},
  \bibinfo{author}{\bibfnamefont{R.~V.~F.} \bibnamefont{Janssens}},
  \bibinfo{author}{\bibfnamefont{D.~G.} \bibnamefont{Jenkins}},
  \bibinfo{author}{\bibfnamefont{T.~L.} \bibnamefont{Khoo}},
  \bibinfo{author}{\bibfnamefont{F.~G.} \bibnamefont{Kondev}},
  \bibinfo{author}{\bibfnamefont{A.}~\bibnamefont{Lopez-Martens}},
  \bibnamefont{\emph{et~al.}}, \bibinfo{journal}{Phys.\ Rev.\ Lett.}
  {\bibinfo{volume}{88}} (\bibinfo{year}{2002}) \bibinfo{pages}{042501}.

\bibitem{willsau93a}
\bibinfo{author}{\bibfnamefont{P.}~\bibnamefont{Willsau}},
  \bibinfo{author}{\bibfnamefont{H.}~\bibnamefont{H{\"u}bel}},
  \bibinfo{author}{\bibfnamefont{W.}~\bibnamefont{Korten}},
  \bibinfo{author}{\bibfnamefont{F.}~\bibnamefont{Azaiez}},
  \bibinfo{author}{\bibfnamefont{M.~A.} \bibnamefont{Deleplanque}},
  \bibinfo{author}{\bibfnamefont{R.~M.} \bibnamefont{Diamond}},
  \bibinfo{author}{\bibfnamefont{A.~O.} \bibnamefont{Macchiavelli}},
  \bibinfo{author}{\bibfnamefont{F.~S.} \bibnamefont{Stephens}},
  \bibinfo{author}{\bibfnamefont{H.}~\bibnamefont{Kluge}},
  \bibinfo{author}{\bibfnamefont{F.}~\bibnamefont{Hannachi}},
  \bibnamefont{\emph{et~al.}}, \bibinfo{journal}{Z. Phys.\ A}
  {\bibinfo{volume}{344}} (\bibinfo{year}{1993}) \bibinfo{pages}{351}.

\bibitem{lopezmartens96a}
\bibinfo{author}{\bibfnamefont{A.}~\bibnamefont{Lopez-Martens}},
  \bibinfo{author}{\bibfnamefont{F.}~\bibnamefont{Hannachi}},
  \bibinfo{author}{\bibfnamefont{A.}~\bibnamefont{Korichi}},
  \bibinfo{author}{\bibfnamefont{C.}~\bibnamefont{Sch{\"u}ck}},
  \bibinfo{author}{\bibfnamefont{E.}~\bibnamefont{Geuorguieva}},
  \bibinfo{author}{\bibnamefont{{Ch.\ Vieu}}},
  \bibinfo{author}{\bibfnamefont{B.}~\bibnamefont{Haas}},
  \bibinfo{author}{\bibfnamefont{R.}~\bibnamefont{Lucas}},
  \bibinfo{author}{\bibfnamefont{A.}~\bibnamefont{Astier}},
  \bibinfo{author}{\bibfnamefont{G.}~\bibnamefont{Baldsiefen}},
  \bibnamefont{\emph{et~al.}}, \bibinfo{journal}{Phys.\ Lett.\ B}
  {\bibinfo{volume}{380}} (\bibinfo{year}{1996}) \bibinfo{pages}{18}.

\bibitem{hauschild97a}
\bibinfo{author}{\bibfnamefont{K.}~\bibnamefont{Hauschild}},
  \bibinfo{author}{\bibfnamefont{L.~A.} \bibnamefont{Bernstein}},
  \bibinfo{author}{\bibfnamefont{J.~A.} \bibnamefont{Becker}},
  \bibinfo{author}{\bibfnamefont{D.~E.} \bibnamefont{Archer}},
  \bibinfo{author}{\bibfnamefont{R.~W.} \bibnamefont{Bauer}},
  \bibinfo{author}{\bibfnamefont{D.~P.} \bibnamefont{McNabb}},
  \bibinfo{author}{\bibfnamefont{J.~A.} \bibnamefont{Cizewski}},
  \bibinfo{author}{\bibfnamefont{K.-Y.} \bibnamefont{Ding}},
  \bibinfo{author}{\bibfnamefont{W.}~\bibnamefont{Younes}},
  \bibinfo{author}{\bibfnamefont{R.}~\bibnamefont{Kr{\"u}cken}},
  \bibnamefont{\emph{et~al.}}, \bibinfo{journal}{Phys.\ Rev.\ C}
  {\bibinfo{volume}{55}} (\bibinfo{year}{1997}) \bibinfo{pages}{2819}.

\bibitem{kruecken96a}
\bibinfo{author}{\bibfnamefont{R.}~\bibnamefont{Kr{\"u}cken}},
  \bibinfo{author}{\bibfnamefont{A.}~\bibnamefont{Dewald}},
  \bibinfo{author}{\bibfnamefont{P.}~\bibnamefont{von Brentano}},
  \bibinfo{author}{\bibfnamefont{D.}~\bibnamefont{Bazzacco}},
  \bibnamefont{and}
  \bibinfo{author}{\bibfnamefont{C.}~\bibnamefont{Rossi-Alvarez}},
  \bibinfo{journal}{Phys.\ Rev.\ C} {\bibinfo{volume}{54}}
  (\bibinfo{year}{1996}) \bibinfo{pages}{1182}.

\end{thebibliography}

\end{document}